\documentclass[manuscript,nonacm]{acmart}

\makeatletter
\def\@ACM@checkaffil{
    \if@ACM@instpresent\else
    \ClassWarningNoLine{\@classname}{No institution present for an affiliation}%
    \fi
    \if@ACM@citypresent\else
    \ClassWarningNoLine{\@classname}{No city present for an affiliation}%
    \fi
    \if@ACM@countrypresent\else
        \ClassWarningNoLine{\@classname}{No country present for an affiliation}%
    \fi
}
\makeatother

\makeatletter
\let\@authorsaddresses\@empty
\makeatother
\AtBeginDocument{%
  }

\setcopyright{rightsretained}
\copyrightyear{2025}
\acmYear{2025}
\acmDOI{10.1145/3630106.3658899}

\acmConference[FAccT '25]{The 2025 ACM Conference on Fairness, Accountability, and Transparency}{June 3--6, 2024}{Rio de Janeiro, Brazil}

\acmBooktitle{The 2024 ACM Conference on Fairness, Accountability, and Transparency (FAccT '24), June 3--6, 2024, Rio de Janeiro, Brazil}

\acmISBN{979-8-4007-0450-5/24/06}

\usepackage{tikz,pgf,wrapfig, xcolor, enumitem}
\usetikzlibrary {arrows.meta}
\usetikzlibrary{backgrounds}
\definecolor{sepia}{HTML}{671800}
\definecolor{light-gray}{gray}{0.01}
\usepackage{xfrac}
\usepackage{stringstrings}
\usepackage{titlesec}
\usepackage{colortbl}  
\usepackage{xcolor}    
\usepackage{booktabs}  
\usepackage{soul}

\NewDocumentCommand{\yi}
{ mO{} }{\textcolor{red}{\textsuperscript{\textit{May}}\textsf{\textbf{\small[#1]}}}}

\begin{document}

\title[Quantitative Insights into Large Language Model Usage and Trust in Academia]{Quantitative Insights into Large Language Model Usage and Trust in Academia:\\An Empirical Study}

\settopmatter{authorsperrow=3} 
\newcommand{\tsc}[1]{\textsuperscript{#1}} 

\settopmatter{authorsperrow=3} 

\author{Minseok Jung}
\email{msjung@mit.edu}
\affiliation{%
  \institution{Massachusetts Institute of Technology}
  \city{Cambridge}
  \state{MA}
  \country{USA}
}

\author{Aurora Zhang}
\email{aszhang@mit.edu}
\affiliation{%
  \institution{Massachusetts Institute of Technology}
  \city{Cambridge}
  \state{MA}
  \country{USA}
}

\author{May Fung}
\email{yrfung@cse.ust.hk}
\affiliation{%
  \institution{Massachusetts Institute of Technology}
  \city{Cambridge}
  \state{MA}
  \country{USA}
}

\author{Junho Lee}
\email{junho.lee@un.org}
\affiliation{%
  \institution{United Nations}
  \city{New York}
  \state{NY}
  \country{USA}
}

\author{Paul Pu Liang}
\email{ppliang@mit.edu}
\affiliation{%
  \institution{Massachusetts Institute of Technology}
  \city{Cambridge}
  \state{MA}
  \country{USA}
}


\begin{abstract}
Large Language Models (LLMs) are transforming writing, reading, teaching, and knowledge retrieval in many academic fields. However, concerns regarding their misuse and erroneous outputs have led to varying degrees of trust in LLMs within academic communities. In response, various academic organizations have proposed and adopted policies regulating their usage. However, these policies are not based on substantial quantitative evidence because there is no data about use patterns and user opinion. Consequently, there is a pressing need to accurately quantify their usage, user trust in outputs, and concerns about key issues to prioritize in deployment. This study addresses these gaps through a quantitative user study of LLM usage and trust in academic research and education. Specifically, our study surveyed 125 individuals at a private R1 research university regarding their usage of LLMs, their trust in LLM outputs, and key issues to prioritize for robust usage in academia. Our findings reveal: (1) widespread adoption of LLMs, with 75\% of respondents actively using them; (2) a significant positive correlation between trust and adoption, as well as between engagement and trust; and (3) that fact-checking is the most critical concern. These findings suggest a need for policies that address pervasive usage, prioritize fact-checking mechanisms, and accurately calibrate user trust levels as they engage with these models. These strategies can help balance innovation with accountability and help integrate LLMs into the academic environment effectively and reliably.

\end{abstract}




\maketitle

\section{Introduction}
Large Language models (LLMs) have significantly enhanced users' ability to retrieve and generate knowledge. These advancements have become particularly prevalent in academic communities \cite{owens2023nature,kasneci2023chatgptedu}, where students, faculty, and researchers often leverage these tools to retrieve and process information. For example, these models are used for brainstorming \cite{owens2023nature}, translation \cite{cabrero2023perceived}, and knowledge acquisition \cite{jo2023decoding}. Despite these opportunities, the reliability of LLMs remains uncertain due to their stochastic nature \cite{bender2021dangers}, and hallucinations and biases from these models often raise questions regarding the trustworthiness of generated text \cite{ji2023hallucination,liang2021towards}.

In response to these concerns, academic institutions have proposed and codified restrictive policies governing the use of LLMs in research and education \cite{ucf_ai_2023}, addressing issues such as information quality \cite{utoronto_ai_classroom_2023}, plagiarism \cite{ucsb_full_ai_writing_policy_2023}, independent reasoning \cite{ucd_ai_and_stuent_writing_2024}, and transparent disclosure \cite{utk_ai_syllabus_statements_2024}. However, current AI policies are typically not based on quantitative evidence \cite{pencheva2020big}, and there is limited evidence that academic AI policies are based on actual LLM usage statistics \cite{utoronto_ai_classroom_2023, Gregory_2023}.
Consequently, there is a demand for real-world studies that explore how usage patterns, perceptions of trust, and critical issues surrounding LLMs are shaped, especially in academia. Understanding the factors influencing usage and trust is crucial for crafting appropriate LLM policies that maximize benefits and minimize harm in academic settings. Additionally, identifying the most critical issues inhibiting usage and trust can also inform the development of LLM research toward addressing these issues.

This paper addresses the aforementioned gaps through a comprehensive study of the usage and trust associated with LLMs. We adopt a data-driven approach, centering on user opinions and experiences to guide the creation of effective and trustworthy LLM practices. Specifically, we ask the following research questions:
\begin{enumerate}
    \item \textbf{RQ1}: \textit{How often do people use LLMs? Are the majority of people using LLMs?}
    \item \textbf{RQ2}: \textit{How are levels of trust in LLM different between those who use the models and those who do not use them?} 
    \item \textbf{RQ3}: \textit{Does increased usage of LLMs positively correlate with higher levels of trust?}  
    \item \textbf{RQ4}: \textit{What are the most important issues inhibiting their trust in LLMs?}
\end{enumerate}

To answer these research questions, we executed an in-depth survey of 125 researchers at a private R1 research university in the United States during the Spring of 2024. Respondents were asked about their average weekly use of LLMs, trust levels, and use purposes. 
They were also prompted to share their concerns surrounding the reliability of LLMs. Based on our studies, we summarize four main findings:
\begin{enumerate}[label=(\arabic*)]
    \item \textbf{Extensive adoption of LLMs}: The results confirm that a substantial portion of people in academic environments actively use LLMs.  
    \item \textbf{Significant relationship between user trust and adoption}:
    There is a significant positive correlation between the perceived trustworthiness of LLMs and users’ decisions to adopt them. In other words, people who regard LLMs as reliable are more likely to incorporate these tools into their daily workflows.
    \item \textbf{Significant relationship between user engagement and trust in LLMs}: More extensive usage of LLMs correlates with greater trust in their outputs, suggesting that hands-on interaction and longer engagement can reinforce user trust. 
    \item \textbf{Critical need to address misinformation}: There is a pronounced demand for fact-checking policies and mechanisms to improve the validation of generated information.
\end{enumerate}

Given these findings, we propose developing LLM policies that both acknowledge academia’s extensive reliance on these tools and prioritize fact-checking protocols to counter misinformation. By adopting a data-driven approach, institutions can ensure that policies remain responsive to evolving needs and accurately account for user behavior. In parallel, LLM companies should strengthen user trust by promoting adoption and fostering deeper engagement. Through this blend of structured oversight and active collaboration, LLMs can continue to mature into reliable and accountable instruments in academic settings.

\section{Related Works}\label{sec:relatedWorks}
\subsection{Deployment of large language models in academic environment}

Various academic institutions are incorporating LLMs into their research and educational practices, each with their own set of rules and points of emphasis. For instance, the University of Massachusetts maintains a positive outlook on using generative LLM for discovering new information \cite{umass_ai_impact_2023}. The University of Central Florida highlights the importance of AI literacy to ensure the effective utilization of these technologies \cite{ucf_ai_2023}. The University of Toronto emphasizes the importance of transformative learning, which involves significantly altering one's perception of technology and society, to effectively integrate LLMs into education \cite{utoronto_ai_classroom_2023}. 

However, there are concerns about LLM dysfunctions such as hallucination, and in response academic institutions have adopted various strategies to address challenges arising from LLMs. For instance, UC Davis underscores the importance of balancing the benefits and risks of AI writing tools \cite{ucd_ai_and_stuent_writing_2024}. Boston University has implemented policies that require the disclosure of any AI assistance used in academic work, emphasizing transparency and academic honesty \cite{bu_gaia_policy_2023}. The University of Massachusetts, Amherst focuses on the appropriate use of LLMs in coursework and research for integrity \cite{umass_ai_impact_2023}. UC Berkeley prioritizes mitigating misinformation from LLMs \cite{berkeley_ai_writing_tools_2023}. Meanwhile, UC Santa Barbara encourages the use of AI tools to support learning \cite{ucsb_full_ai_writing_policy_2023}. 

Other academic communities have adhered to more traditional principles, with prominent journals such as \textit{Nature} and \textit{Science} regulating AI usage in research \cite{nature2024aipolicies, science2024aipolicies}. Major conferences in computer science like the International Conference on Machine Learning (ICML) \cite{icml2024aipolicies}, Association for the Advancement of Artificial Intelligence conference (AAAI) \cite{aaai2024aipolicies}, and the Conference on Neural Information Processing Systems (NeurIPS) \cite{neurips2024aipolicies} have focused on the ethical implications, misinformation, and regulatory aspects of LLM, reflecting the growing consensus of AI governance.



\subsection{Trust in LLMs}
There is extensive literature documenting the importance of trust in user experiences of automated systems. Related work has defined and operationalized the concept of trust through various frameworks and metrics~\cite{mcknightTrustSpecificTechnology2011a,korberTheoreticalConsiderationsDevelopment2019,chiouTrustingAutomationDesigning2023, gulatiDesignDevelopmentEvaluation2019a, kohnMeasurementTrustAutomation2021}. Some have pointed out the multidimensional nature of trust and the multiplicity of definitions that it may entail. Others define ``trust'' as a positive belief in another party’s goodwill, a cognitive state justified by reasons or arguments, or an emotional state that depends on one’s feelings towards another party \cite{leeTrustAutomationDesigning2004, uenoTrustHumanAIInteraction2022}.

Quantitative methods can provide insight into how user trust may affect their experiences with LLM. Bahmanziari et al. use policy capturing, a simulation-based model, to study the relationship between trust and technology adoption \cite{bahmanziariTrustImportantTechnology2003}. Several quantitative studies have also been conducted around trust in language models. Kim et al. use regression analysis to show that LLMs that express uncertainty reduce overreliance on AI but may also reduce trust overall \cite{kimNotSureExamining2024}. Cabrero-Daniel and Cabrero conduct a quantitative study on user expertise and trust in natural language generators \cite{cabrero-danielPerceivedTrustworthinessNatural2023}. Finally, related work has used structural equation modeling as a method: Jo investigates the different factors that lead to language model adoption, including trust, \cite{joDecodingChatGPTMystery2023} and Choudhury and Shamszare study the correlation between trust and intent to use ChatGPT in a medical setting \cite{choudhuryInvestigatingImpactUser2023}.
Part of the literature has also studied how to improve trust in AI. Trust may be fostered by aligning algorithmic development around desirable values such as accuracy, explainability, and bias-awareness \cite{srinivasanImprovingTrustData2020, rossiBuildingTrustArtificial2018}. Other studies have shown that it might be helpful to use tools such as interactive visualization or even ChatGPT to improve trust in AI in certain contexts \cite{beauxis-aussaletRoleInteractiveVisualization2021, yeImprovedTrustHumanRobot2023}.

This study contributes to the existing literature by employing quantitative methods to examine trust and the use of LLMs in an academic setting. Furthermore, this study provides interpretations of the findings to improve the understanding of how to build trust responsibly in LLMs, with a particular emphasis on how this contextualized knowledge can inform the development of AI policies in academic institutions.

\subsection{Major issues facing LLM deployment}\label{major_issues}
Academic institutions face a variety of issues related to LLM policies, including trustworthiness \cite{ucf_ai_2023}, potential misinformation \cite{utoronto_ai_classroom_2023}, plagiarism \cite{ucsb_full_ai_writing_policy_2023}, delegation to LLM tools \cite{ucd_ai_and_stuent_writing_2024}, and transparency \cite{utk_ai_syllabus_statements_2024}. At this time, there is no consensus on a single issue to prioritize for LLM policy. While some institutions prioritize ex-post measures such as disclosing the tools and penalty for a violation of integrity ~\cite{ucsb_full_ai_writing_policy_2023,upitt2023,utk_ai_syllabus_statements_2024}, others focus on ex-ante strategies aimed at ensuring ethical usage and accurate information~\cite{berkeley_ai_writing_tools_2023,utoronto_ai_classroom_2023}. This diversity reflects each institution's differing priorities and concerns, highlighting the need for a more unified approach to prioritize and address the most critical issues around LLMs.

\section{Methodology}\label{sec:methodology}
\subsection{Research Questions}\label{questions}

As LLMs become more integrated into academic settings, it is crucial to understand the factors related to usage and trust in LLMs. In this research, we define \textit{LLMs} as AI architectures designed to process, interpret, and generate human-like text through deep learning techniques, particularly neural networks~\cite{openai2024chatgpt}. We employ the term \textit{academia} to refer to the collective ecosystem of higher education institutions, research organizations, and scholarly communities engaged in the pursuit of knowledge production, dissemination, and education~\cite{altbach2019trends}.
Our research focuses on four key questions: 
\begin{enumerate}
    \item \textbf{RQ1}: \textit{How often do people use LLMs? Are the majority of people using LLMs?}
    \item \textbf{RQ2}: \textit{How are levels of trust in LLM different between those who use the models and those who do not use them?} 
    \item \textbf{RQ3}: \textit{Does increased usage of LLMs positively correlate with higher levels of trust?}  
    \item \textbf{RQ4}: \textit{What are the most important issues inhibiting their trust in LLMs?}
\end{enumerate}

\begin{figure}[t!]
    \centering
    \includegraphics[width=3.3in]{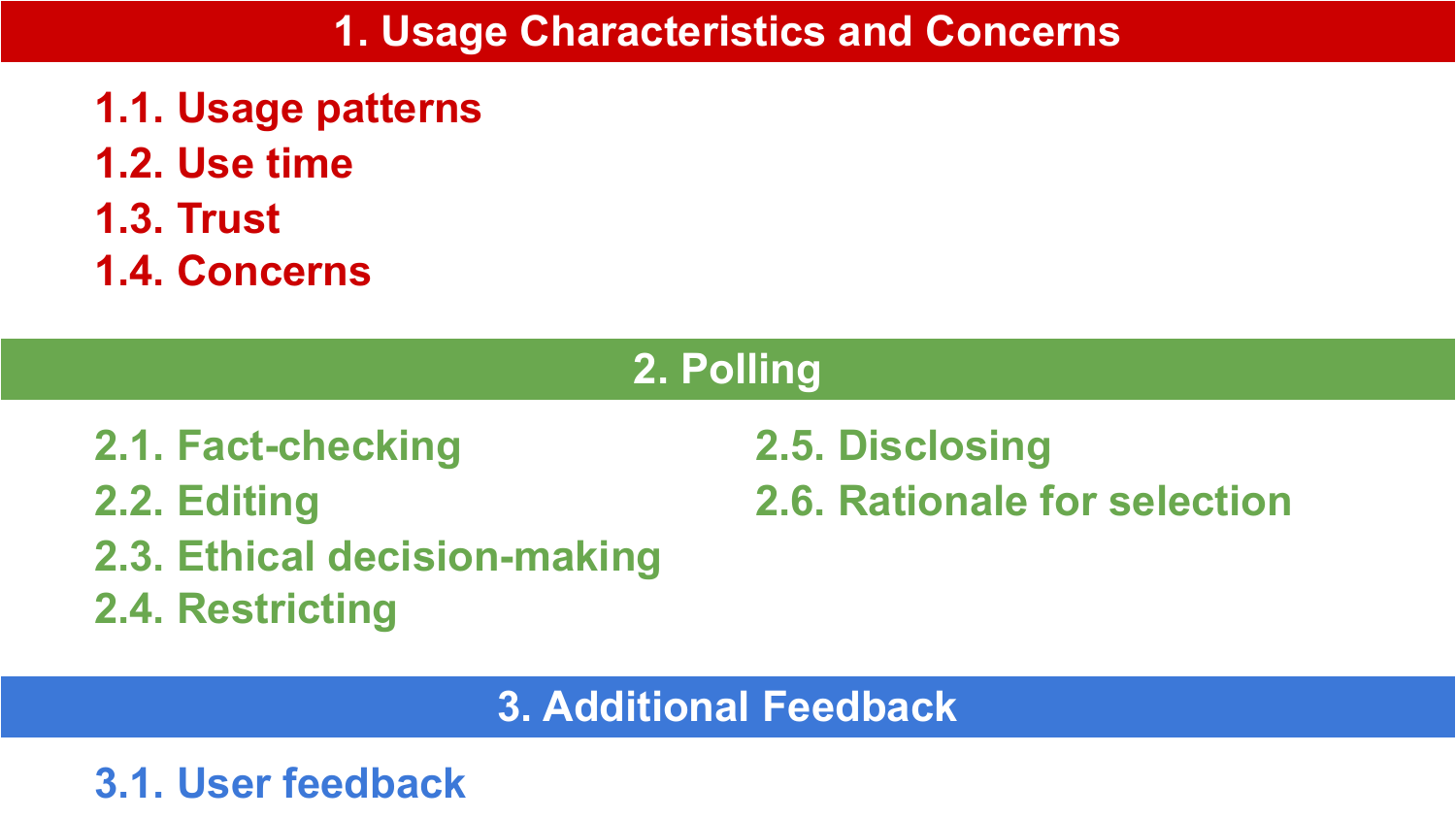}
    \caption{An overview of the primary topics examined in our user study on language model usage in academia. The figure highlights two main phases as well as additional feedback: usage characteristics and concerns (red), polling to five core issues (green), and additional feedback (blue). The first section (red) answers \textbf{RQ1-3} and the second section answers \textbf{RQ4}. The additional feedback was included for further development.}
    \label{fig:flowchart}
\end{figure}

\textbf{RQ1} is fundamental in gauging the prevalence and reach of LLMs in everyday life, providing a baseline for further analysis. \textbf{RQ2} helps us understand the trust dynamics, which is related to barriers to adoption and areas needing improvement. \textbf{RQ3} seeks to determine whether engagement breeds trust, which has significant implications for use patterns and the design of more trustworthy systems. \textbf{RQ4} identifies the primary concerns of users to prioritize further features and safeguards that align with user needs and societal values. These five major concerns -- fact-checking, editing, ethical decision-making, restricting, or disclosing -- are shortlisted through a thorough thematic analysis of AI policies and guidelines from academic institutions~\cite{umass_ai_impact_2023,ucsb_full_ai_writing_policy_2023, ucd_ai_and_stuent_writing_2024,ucf_ai_2023,umass_ai_impact_2023,utoronto_ai_classroom_2023,berkeley_ai_writing_tools_2023,bu_gaia_policy_2023}, and are defined as follows: \textit{fact-checking}: verifying the accuracy of generated content, \textit{editing}: modifying LLM outputs to create their own content, \textit{ethical decision-making}: avoiding misuse of LLM outputs based on contextual considerations, \textit{restricting}: determining the conditions under which LLMs are permitted for use in specific situations, and \textit{disclosing}: acknowledging the use of AI tools.

\subsection{Experimental Design}
The overview of the survey is in Figure \ref{fig:flowchart}. The survey contained a structured set of questions divided into multiple sections, with approximately 20-25 questions and conditional questions based on the participant's responses. The prompts included questions about the participant's role at the private school, their usage of language models, their opinion on the importance of understanding the technical process behind LLMs, and their trust in AI-generated content. We focused on five key features: LLM adoption (binary), usage duration (ordinal), trust levels (ordinal), importance scores of each issue (ordinal), and qualitative comments on each issue (written text). Ordinal data was collected using a five-point Likert scale. The survey was also designed to study participants' attitudes toward five key topics aforementioned in \textit{RQ4}. Each participant was asked to cast a single vote per topic, rating them on a scale of 1 to 5, with a larger number indicating higher importance. Alongside their ratings, participants provided written rationales for their choices. These qualitative comments give deeper insights into these topics, offering a nuanced understanding of the concerns that matter most to the academic community.


We intentionally excluded demographic features such as age and gender to reduce participants' fatigue, aiming for a completion time of under 10 minutes. The pre-testing was conducted a week before the approved survey period, where we received feedback from experts in Computer Science and Statistics outside the target population to check the question prompts. The experiment was conducted around a month as described in section \ref{approval}. The high completion rate of 88\% indicates that the survey was effectively designed to minimize fatigue.



\subsection{Approval}\label{approval}
This study has received approval from the institutional review board (IRB) of the selected school, with a petition for approval submitted under the title of ``Decoding AI: A Comprehensive Survey on Language Model Usage, Perceptions, and Policies at [institution name].'' The experiment was exempted from further institutional regulation due to its exclusion of sensitive demographic features such as race, ethnicity, and financial background. The initial allocated period for data collection spanned from April 8 to April 22, 2024, but in response to significant interest and formal petitions for an extension, the IRB sanctioned an extension of the study period to May 6, 2024.

\subsection{Study population}
The study's target population includes students, research staff, and faculty at the selected private academic institution. This population was selected as the target to assess the use of LLM tools in academic settings. Non-academic staff was not included, and membership was checked by the institutional email address.

The use of a relatively localized sample can yield meaningful insights that extend beyond the given domain. Researchers often conduct studies within a single institution to explore phenomena that can be generalized to broader populations. While the restriction in source and time may vary and there could be special needs for a specific research topic (e.g. the project is funded for an institutional study), results from local samples still offer legitimate grounds for broader inferences. For instance, \citet{bahmanziari2003trust} investigate trust and technology adoption decisions based on 181 responses from a U.S. business school, demonstrating how individual organizations can serve as microcosms for broader phenomena. Similarly, \citet{zadeh2019social} generalize findings from 50 annotators at a research university to judge the social intelligence of AI systems. Other studies employ a similarly structured approach but focus on different populations. For example,~\citet{jo2023decoding} examines chatbot adoption in South Korean colleges, extrapolating a finding from a country to the broader literature. Beyond individual institutions or countries,~\citet{cabrero2023perceived} recruit participants ($n=77$) via LinkedIn to investigate the perceived trustworthiness of generative text models, while~\citet{choudhury2023investigating} survey a larger, web-based cohort ($n=607$) on ChatGPT usage and trust. 

While the single-institution sample presents certain limitations, prior research has shown that a local setting can function as microcosms of global phenomena. Thus, examining participants from one institution can still yield credible insights that extend beyond the immediate context.


\subsection{Data collection method}
This study recruited participants using Qualtrics~\cite{qualtrics2024}, selected for its robust data security features, ensuring the protection of participant information and the integrity of the data collected. Recruitment and survey distribution were conducted via email, targeting multiple laboratories and research communities while ensuring a diverse representation of opinions from various roles and departments within the institution.

\subsection{Data pre-processing}
A total of 88 samples were secured after conducting data pre-processing from 125 instances. Along with validation using partitioning~\cite{kitchenham2008personal}, a rigorous filtering process was conducted to ensure the integrity and relevance of the dataset. First, data entries outside the expected data collection period were excluded, reducing the number of responses from 125 to 96. Subsequently, only participants who provided explicit consent for data usage were included in the dataset, though this step did not lead to a further reduction. The process then filtered out roles outside the target population — the institution's students, research staff, and faculty —  and entries that were not affiliated with any of the six schools of the institution to maintain contextual relevance, resulting in a reduction to 90 responses. Finally, email verification was conducted, ensuring the validity of responses, which led to a final dataset of 88 valid responses. The dataset encompasses a participation breakdown of 76 students and 12 non-students, with the latter comprising 7 research staff and 5 faculty members. The details about the roles are noted in Appendix \ref{fig:roles_school}.

\subsection{Data Analysis}
To extract meaningful insights from user patterns, trust levels, and the prioritization of issues related to LLMs, this study employed four main data analysis techniques. Each method was selected to address the corresponding research question effectively, based on its suitability for the data types involved. Detailed methods are noted in Appendix \ref{appendix:quantatitive_analysis}.

\subsubsection{Frequency Analysis.}
To address \textbf{RQ1}, we conducted a frequency analysis to observe the distribution of responses regarding LLM usage. We assessed the frequency of use with five bins corresponding to the five levels of the Likert scale.

\subsubsection{Point-Biserial Correlation Coefficient.}
For \textbf{RQ2}, we investigated the relationship between LLM adoption and trust levels. Since LLM adoption is a binary variable (non-users vs.\ users) and trust levels are ordinal data, the point-biserial correlation coefficient was deemed the most appropriate measure. This method quantifies the association between a binary variable and an ordinal variable, allowing us to measure the difference between binary features. The detailed equation of the point-biserial correlation coefficient and statistical significance testing is provided in Appendix~\ref{appendix:point-biserial}. We explain how it measures the mean difference in trust levels between the two groups relative to the overall variability in trust scores. 


\subsubsection{Kendall's Tau.}
Addressing \textbf{RQ3}, we analyzed the ordinal association between self-reported use time and trust levels. Both variables are ordinal and may not meet the assumptions of parametric tests (such as normality and linearity). Therefore, Kendall's tau, a non-parametric measure of rank correlation, was selected as the most suitable method. Kendall's tau assesses the strength and direction of association between two ordinal variables. It accounts for concordant and discordant pairs of observations. The mathematical details of Kendall's tau calculation are in Appendix~\ref{appendix:kendall-tau}. This method is ideal for our analysis as it captures the monotonic relationship between usage duration and trust levels. Also, to visualize the relationship and account for non-linear associations, we employed the locally weighted scatterplot smoothing (LOWESS) regression. This method fits a smooth curve to the data for trends that may not be captured by linear models.

\subsubsection{Ranking.}
For \textbf{RQ4}, we aimed to identify which of the shortlisted five core issues—\textit{fact checking, editing, ethical decision-making, restricting, and disclosing} is perceived as most important by LLM users. We utilized a score-based ranking method derived from participants' ratings on a Likert scale of 1 to 5 for each issue. This method is effective because it aggregates individual ratings into an overall ranking, highlighting the issues that are most salient to the academic community. The ranking method is noted in Appendix~\ref{appendix:ranking}.

\subsubsection{Qualitative Analysis.}
Finally, to gain deeper insights into participants' perspectives and reasoning, a qualitative analysis of users' written rationales was conducted. The authors engaged in a two-month consultation with experts in user studies and IT policy research to thoroughly review and interpret the responses. The qualitative methods are noted in Appendix~\ref{appendix:qualitative_analysis} along with summaries noted in Appendix~\ref{appendix:opinion_summaries}. This process ensured that the participants' views on each issue were captured and sample opinions are in Table~\ref{opinion_table}.

\section{Results}\label{sec:results}

The following findings directly correspond to each inquiry. Finding 1 relates to \textbf{RQ1}, demonstrating that a significant majority of individuals utilize language models, thereby confirming the widespread adoption of LLMs in academic settings. Finding 2 and finding 3 pertain to \textbf{RQ2} and \textbf{RQ3}, respectively, by revealing a positive correlation between LLM adoption and trust levels, as well as showing that increased engagement with LLMs is associated with higher trust with statistical significance in both cases. Finally, finding 4 addresses \textbf{RQ4} by identifying fact-checking as the most important concern among LLM users, revealing a critical area that needs attention to enhance the reliability and ethical deployment of LLMs. Collectively, these findings provide empirical evidence that directly answers the research questions.

\subsection{Finding 1: Significant portion of people utilize LLMs}\label{finding1}
\begin{figure}[t!]
    \centering
    \includegraphics[width=3.5in]{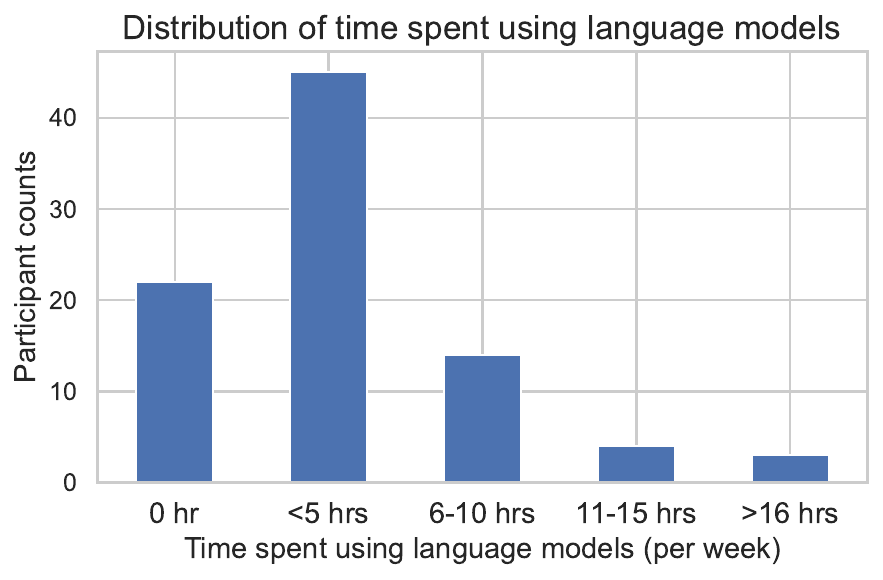}
    \caption{Distribution of respondents' weekly use time of language models. The majority of respondents use LLMs and most use them for between 1 and 5 hours per week. Also, 25\% of respondents replied that they are not using those tools.}
    \label{fig:use_time}
\end{figure}

Although a high volume of LLM adoption was anticipated, there have been limited efforts to quantify the amount of usage. This analysis contributes to the literature by providing new quantitative data that finds that the majority of people use LLMs. Fig. \ref{fig:use_time} shows that 75\% (66 out of 88) of people use LLMs for research, indicating a high level of usage.

The majority of users, ($n=45$), interact with the tools for ``Less than 5 hours,'' indicating a preference for brief usage. In contrast, 22 people replied that they do not utilize the tool at all. A smaller segment, comprising 14 users, engages moderately with the tools for ``6 to 10 hours.'' Fewer users are observed in the higher usage brackets, with only 4 users spending ``11 to 15 hours'' and 3 users exceeding ``16 hours.''

\subsection{Finding 2: Positive correlation between trust level and adoption}\label{finding2}

As the finding that the pervasive adoption of LLMs for academic environments could bring up divergence of the trust to the models, the research aimed to elucidate insights into the different trust levels between people who use LLMs and people who do not. The analysis was conducted with the hypothesis that \textit{individuals who use language models exhibit higher levels of trust in these models compared to those who do not use them.}

\begin{figure}[hbt!]
    \centering
    \includegraphics[width=3.5in]{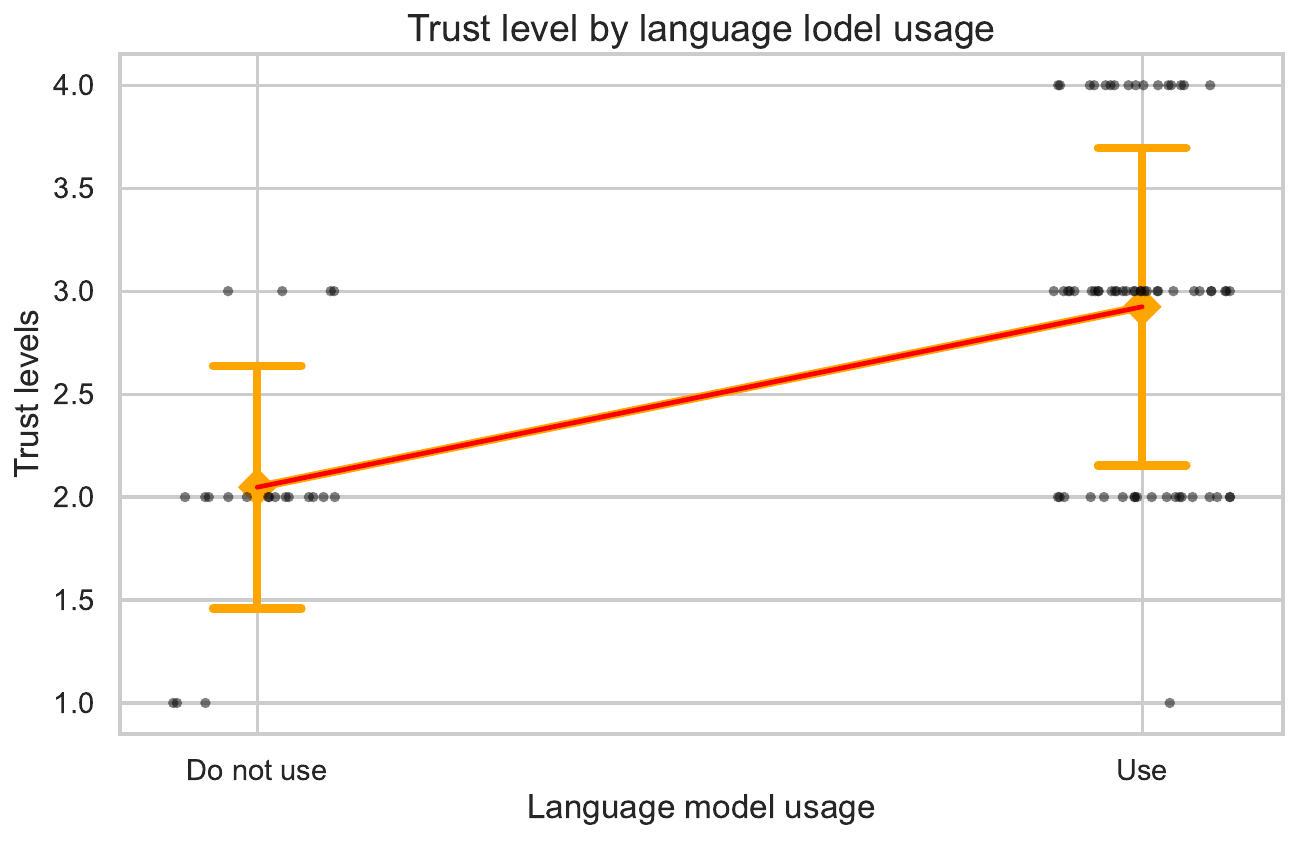}
    \caption{Point-biserial correlation shows the relationship between language model usage (``Do not use'' vs. ``Use'') and trust levels on a Likert scale. Each dot represents the number of respondents at trust levels 1 (``Completely distrust'') to 4 (``Mostly trust''), with no responses for level 5 (``Completely trust''). Each dot indicates a reply, and the middle point is the average of each group. The cap indicates the extent of the error bar, which is about the standard deviation above and below the mean. The line connecting mean values for ``Do not use'' and ``Use'' groups shows a moderate positive correlation (r = 0.4601), indicating adoption links to higher trust. This correlation is statistically significant (p = 7.316e-06).}
    \label{fig:correlation1}
\end{figure}

This study shows a moderate positive correlation between adoption and trust as shown in Figure~\ref{fig:correlation1}. We also include the mean and standard deviation for the trust levels within each group ``Do not use'' and ``Use'', showing diverging trust levels of language models based on usage. The point-biserial correlation coefficient of 0.4601, with a highly significant p-value (7.316e-06), indicates a moderate positive correlation between adopting LLMs and higher trust levels. Users of language models exhibit a higher average trust level (around 3; ``neutral'') compared to non-users (around 2; ``mostly distrust''). This suggests that the adoption of LLMs is associated with increased trust in these tools, potentially due to familiarity and perceived reliability developed through usage.

\subsection{Finding 3: Positive correlation between trust level and engagement} \label{finding3}

\begin{figure*}
  \centering
  \includegraphics[width=1.0\textwidth]{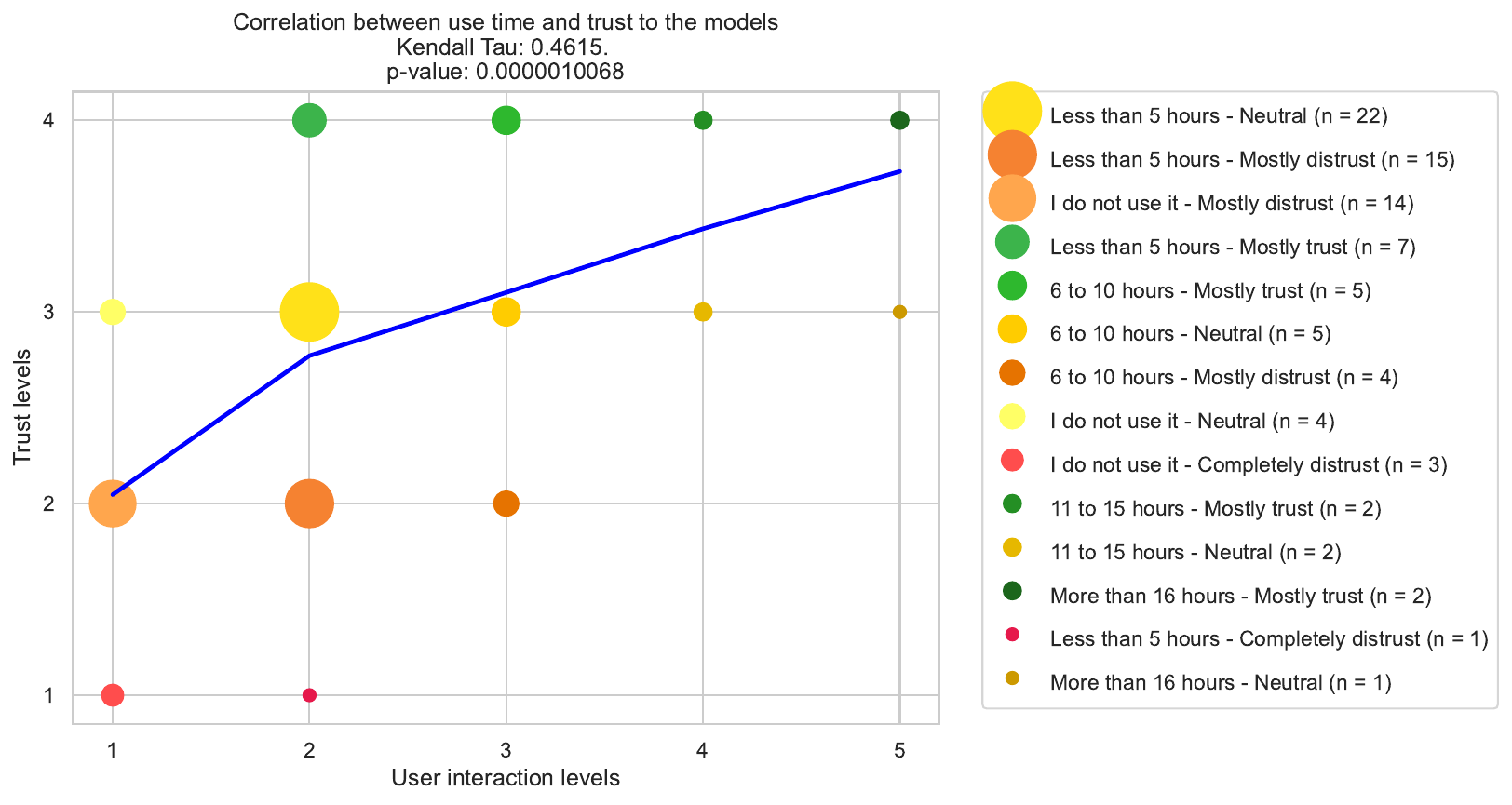}
  \caption{A positive moderate correlation coefficient between trust levels and use time is monitored using Kendall’s tau (with a Lowess fit represented by the blue curve). The color indicates different trust levels, with red representing `Completely distrust', orange for `Mostly distrust', yellow for `Neutral', and light green for `Mostly trust'. No responses indicated `Completely trust'. The size of the dots represents the number of respondents, and color intensity increases with the amount of use time. The result indicates that users who spend more time using language models tend to have higher trust levels in them.}
  \label{fig:correlation2}
\end{figure*}

An in-depth analysis was conducted based on the hypothesis: \textit{increased engagement of language models correlates with higher levels of trust in these models.} Figure~\ref{fig:correlation2} displays the correlation between the time spent using LLMs and the users' trust levels.  

The correlation is quantified using Kendall's tau, which is 0.4615, indicating a moderate positive relationship. This suggests that as users spend more time using the tools, their trust in the LLMs generally increases. The statistical significance of this correlation is supported by a p-value of 1.068e-06, demonstrating that the observed relationship is highly significant. In Fig \ref{fig:correlation2}, the trust level of 5 (``Completely trust'') is not plotted because no one answered they completely trust the LLMs.

\subsection{Finding 4: Fact-checking is perceived to be the most important issue} \label{finding4}
The five issues we examine in our study, fact-checking, editing, ethical decision-making, restricting, and disclosing, are significant in the context of LLM usage because they directly impact the reliability, integrity, and ethical implications of AI-generated content. Fact-checking ensures the accuracy and credibility of information generated by LLMs, mitigating the risk of disseminating false or misleading information. Editing refines the clarity and coherence of LLM outputs, making them more useful and relevant. Ethical decision-making is paramount in guiding how LLMs are used, ensuring that they align with societal values and norms. Restricting access and use of AI tools can prevent misuse and protect sensitive information, while disclosing the use of LLMs maintains transparency and accountability, fostering trust among users. To identify the most important issue among them, we asked users to vote on their significance, and Table \ref{tab:voting} summarizes the result.

\begin{table}[t]
    \setlength\tabcolsep{7.0pt}
    \centering
    \vspace{-0mm}
    \begin{tabular}{lccccc}
    \toprule
    Rating & Check & Edit & Decide & Restrict & Disclose \\
    \midrule
    5 & \cellcolor{green!50}\textbf{37.35\%} & \cellcolor{green!50}20.73\% & \cellcolor{green!50}29.11\% & \cellcolor{green!50}18.39\% & \cellcolor{green!50}25.97\% \\
    4 & \cellcolor{green!40}\textbf{43.37\%} & \cellcolor{green!40}42.68\% & \cellcolor{green!40}41.77\% & \cellcolor{green!40}41.38\% & \cellcolor{green!40}35.06\% \\
    3 & \cellcolor{gray!30}10.84\% & \cellcolor{gray!30}26.83\% & \cellcolor{gray!30}21.52\% & \cellcolor{gray!30}\textbf{33.33\%} & \cellcolor{gray!30}28.57\% \\
    2 & \cellcolor{red!20}6.02\% & \cellcolor{red!20}\textbf{7.32\%} & \cellcolor{red!20}3.80\% & \cellcolor{red!20}4.60\% & \cellcolor{red!20}6.49\% \\
    1 & \cellcolor{red!10}2.41\% & \cellcolor{red!10}2.44\% & \cellcolor{red!10}3.80\% & \cellcolor{red!10}2.30\% & \cellcolor{red!10}\textbf{3.90\%} \\
    \bottomrule
    \end{tabular}
    \vspace{1mm}
    \caption{Value percentile table for importance ratings across different categories: (5-Strongly Important, 4-Important, 3-Neutral, 2-Unimportant, 1-Strongly Unimportant), and (Check: Fact-checking, Edit: Editing, Decide: Ethical Decision-making, Restrict: Restricting, and Disclose: Disclosing). The results indicate that fact-checking is perceived as the most important issue among language model users. The color gradients represent the level of importance, with green indicating higher importance, gray denoting neutral importance, and red signifying lower importance, while bold values highlight the highest percentage within each category.}
    \vspace{-2mm}
    \label{tab:voting}
\end{table}

Among core issues, ``fact-checking'' emerged as the most critical issue, with 37.35\% of respondents rating it as ``Strongly Important'' and 43.37\% as ``Important.'' Editing is also highly regarded, with 20.73\% of respondents considering it ``Strongly Important'' and 42.68\% as ``Important.'' Ethical decision-making follows, with 29.11\% rating it as ``Strongly Important'' and 41.77\% as ``Important.'' Restricting and disclosing LLM use are perceived similarly in importance, with 18.39\% and 25.97\% respectively rating them as ``Strongly Important'' and 41.38\% and 35.06\% as ``Important.'' However, disclosing is seen as less critical compared to the others, as indicated by the lower proportion of ``Strongly Important'' rating of 25.97\%.

In the written rationales for their selections, respondents (denoted as ``P'') stressed the importance of fact-checking to prevent misinformation and to secure scientific integrity, saying \textit{``a strict rule of human verification before publication is essential''} (P76). Editing was seen as crucial for maintaining originality: \textit{``the crux of the argument or fact is the important part that should not be generated''} (P90). Ethical decision-making was considered crucial for human oversight to evaluate potential biases and ensure responsible use as \textit{``Human's role in final decision-making is central''} (P116). Regarding restricting LLM usage, respondents advocated for a partial regulation of usage and noted it is \textit{``Helpful to allow flexibility depending on specific needs in the academic setting. Otherwise, one seems to fool itself by avoiding new technology and will one day be left behind by the new generation''} (P116). Disclosing AI involvement was essential for maintaining transparency: \textit{``If an LLM is used, it needs to be fully and properly documented in the same way that computational science researchers acknowledge their methods, software, etc''} (P60).

In particular, participants emphasized that a mandatory human fact-checking process for LLM-generated content is critical, and noted the need for \textit{``a forcing function that checks the facts as a mandatory step of the publishing process''} (P98). Respondents mentioned that such a process is essential to ensure the accuracy and accountability of published materials, which is particularly important given the prevalence of AI tools. Even though users can use LLMs, the users \textit{``should not blindly trust the output''} (P46), as this could undermine the integrity of scholarly publications and erode trust in academic communities.

Outside of fact-checking, participants expressed opinions about other issues including editing, ethical decision-making, restricting, and disclosing. Table \ref{opinion_table} highlights representative responses for each category and Appendix \ref{appendix:opinion_summaries} includes summaries of the excerpted replies.

\begin{table}
\centering
\renewcommand{\arraystretch}{1.0} 
\setlength{\tabcolsep}{8pt} 
\begin{tabular}{p{1.4cm}p{10.5cm}p{1.5cm}}
\toprule
\textbf{Topic}         & \textbf{Opinion}                                                                                                                                            & \textbf{Participant} \\ \midrule
\textbf{Fact-Checking} & \textit{One student in the class that I work as a TA cheated on the assignment with AI. In academia, I am seriously concerned about plagiarism and losing critical thinking skills.} & P14 \\
                       & \textit{Maintain the highest scientific standards.}                                                                                                         & P31 \\
                       & \textit{Students should not blindly trust the output and do the work to verify.}                 & P46 \\
                       & \textit{A strict rule of human verification before publication is essential.}               & P76 \\
                       & \textit{Avoid discrepancies between AI-generated content and facts.}                                                                                               & P87 \\ \midrule
\textbf{Editing}       & \textit{We risk the utter deterioration of people’s writing ability and a boring world where everything is written to sound exactly the same.}                     & P47 \\
                       & \textit{AI is a useful tool. If you use it and plagiarize or introduce errors, you are responsible for those problems.} & P74 \\
                       & \textit{I’m happy to have LLM check my grammar and make my sentences more coherent.}                                                                               & P82 \\
                       & \textit{There is a trade-off to be made in human supervisory roles vs. time cost into checking everything through the editorial process.}                           & P116 \\ \midrule
\textbf{Ethical Decision} & \textit{We’re releasing and using tools that are already wreaking havoc in society (election and war disinformation, deepfake pornography). All models must be rigorously tested.} & P47 \\
                       & \textit{Evaluate AI tools with the same criteria as we do non-AI works and opinions.}                                                                               & P80 \\
                       & \textit{AI does not know how to set priorities. Humans should design their own society and not let virtual agents tell us what is important.}                      & P82 \\
                       & \textit{Again, it is choosing the middle ground. Banning the tech or highly regulating it is no solution. Just pass it through the same pair of scrutinizing eyes as human text.} & P90 \\ \midrule
\textbf{Restriction}   & \textit{It is important to use wisely and think critically.}                                                                                                       & P12 \\
                       & \textit{Let AI be AI! If students want to have AI write their paper and be creative, why isn’t that a good thing?}                                                & P80 \\
                       & \textit{Research institutions should be open-minded on technical breakthroughs.}                                                                                  & P82 \\
                       & \textit{Helpful to allow flexibility depending on specific needs in the academic setting. Otherwise, one seems to fool oneself for avoiding new technology.}       & P116 \\ \midrule
\textbf{Disclosure}   & \textit{I don’t want my students to use AI for their sources.}                                                                                                     & P12 \\
                       & \textit{If an LLM is used, it needs to be fully and properly documented in the same way computational science researchers acknowledge their methods.}             & P60 \\
                       & \textit{It’s a tool. Human beings find tools that are useful and learn to use them.}                                                                               & P87 \\
                       & \textit{Reproducibility that is central to research should be upheld.}                                                                                            & P116 \\ \bottomrule
\end{tabular}
\caption{A subset of the respondents' opinions on key issues in LLM usage. They are all open-ended questions. Participants exhibited considerable concerns alongside optimistic perspectives regarding the use of LLMs in academic environment.}
\label{opinion_table}
\end{table}

\section{Discussion}\label{sec:discussion}




\subsection{Needs for LLM usage policy in academic environment}
The widespread adoption of LLMs within academic settings, as evidenced by our study where 75\% of respondents actively utilize LLMs for research, underscores the needs for AI policies in academia, especially in activities related to knowledge generation. Our findings reveal that while LLMs enhance knowledge retrieval and generation, concerns about misinformation and academic integrity persist. For instance, the significant emphasis placed on fact-checking highlights the necessity for usage policies that mandate rigorous verification of AI-generated content to maintain scholarly standards. This is especially important in fields that require accurate knowledge, like medical research, where inaccuracies can have serious health implications \cite{monteith2024artificial, choudhuryInvestigatingImpactUser2023}.  

\subsection{Needs for addressing hallucination from LLMs}
One of the paramount challenges identified in our study is the issue of hallucinations. This phenomenon significantly undermines the trustworthiness of LLMs and poses a threat to knowledge dissemination. Academic policies should mandate the implementation of robust fact-checking protocols, ensuring that all AI-generated content undergoes thorough verification before being utilized in scholarly work. Educating users about the limitations of LLMs and promoting best practices for critical evaluation of AI outputs can also play a vital role in mitigating the impact of hallucinations while developing technologies to minimize generation of misinformation. By prioritizing the resolution of hallucination issues, academic institutions can safeguard the quality and credibility of their research outputs.

\subsection{Recommendation for using a data-driven approach in developing AI policy}

A data-driven approach in AI policy development is essential for creating effective and inclusive frameworks reflecting the academic community's needs. It uses real-world user data, grounding policies in data and diverse perspectives, unlike traditional expert-driven methods. Our study’s quantitative insights into usage patterns and trust levels provide a foundational understanding that can inform evidence-based policy-making. Policies based on quantitative methods could be especially convincing for AI users. 

By continuously collecting and analyzing data on how LLMs are utilized, researchers can identify emerging trends, user behaviors, and evolving concerns, allowing for the timely adaptation of policies to meet the needs of the academic community. Furthermore, advanced analytics can be employed to explore complex relationships within the data, uncovering nuanced insights that might remain hidden through simplistic analysis.



\subsection{Increasing the exposure to build trust}
The study shows that exposure to LLMs is crucial in building trust among academic users. LLM companies can enhance their presence in academia by partnering with educational institutions, offering training workshops, and integrating LLMs into both curricula and research. This exposure could help people experience LLMs' capabilities and enhance trust through interaction. For example, showcasing successful case studies of LLM applications in research or education can demonstrate their practical value and reliability. We can also interpret these results more cautiously: As users become more exposed to LLMs, they may trust them more. Therefore, it is crucial to create safeguards and accountability systems to ensure that user trust is not misplaced. As LLMs become increasingly prevalent in academic settings, this points to an increased need for reliable and trustworthy systems that merit users' trust.


\subsection{Use of AI content detector}

AI detectors that effectively flag and filter AI-generated content can have great potential to mitigate the spread of fabricated misinformation. Even though AI content detectors are not completely reliable~\cite{macko2023multitude}, they are one of the most efficient ways to check large volumes of content \cite{mitchell2023detectgpt}. Some academic institutions have already begun employing AI content classifiers to classify AI-generated content. For example, Boston University's Faculty of Computing and Data Sciences encouraged the use of auto-detection tools \cite{bu_gaia_policy_2023} and Texas Tech University's Library recognized them as a useful tool \cite{Gregory_2023}. In contrast, Baylor University expressed doubts about the uncertainty of the detection process \cite{baylor2023} and the University of Pittsburgh prohibited the usage of AI detectors due to their unstable error rates and concerns over unequal treatment of essays by non-native English speakers \cite{upitt2023}.

Looking to the future, the adoption of AI content classifiers could greatly benefit academia by preserving the quality of research and reducing the costs associated with reviewing and verifying work.

\subsection{Limitations and future work}

The study's population is limited to a private research institution, so the results may reflect the distinct characteristics of the school. Considering the school itself is STEM-centric, 83\% of respondents answered that it is important to have an understanding of the technical process behind LLMs as noted in Appendix \ref{appendix:understanding_process}. This trend is detected in the respondents' affiliated department(s) as well. The collected data indicates that 49.5\% (n = 48) of the responses came from respondents affiliated with Engineering disciplines. Computing followed with a participation rate of 25.8\% (n = 25). Contributions from the School of Science accounted for 13.4\% (n = 13), while Management represented 6.2\% (n = 6). Architecture discipline made up 4.1\% (n = 4) of the responses, and fields within Social Sciences contributed one response. Multiple selections were allowed in this survey. 

Since the school itself is significantly technology-oriented, the unique characteristics of private research institutions may not fully represent broader patterns in different types of academic settings. Future research should expand to include a broader range of academic settings, such as liberal arts colleges, public universities, and institutions with a stronger focus on humanities or social sciences. Additionally, disciplinary comparisons could reveal how trust and usage patterns vary between STEM and non-STEM fields could lead to meaningful results. Also, we could conduct experiments based on different environments, including a broader population of general users outside of academia, to provide a more inclusive understanding of LLM usage and trust.

\section{Conclusion}\label{sec:conclusion}
This study investigated the user patterns and trustworthiness of language models in academic settings, revealing widespread adoption of LLMs, with 75\% of respondents actively using them. A significant positive correlation was found between engagement and trust, as well as between adoption and trust, suggesting that increased exposure to LLMs fosters user confidence. Fact-checking emerged as the most critical concern, highlighting the need for robust mechanisms to mitigate misinformation and ensure the reliability of AI-generated content. These findings have profound implications for stakeholders across sectors. For AI companies, fostering trust requires strategies that promote adoption and engagement, such as user-friendly interfaces and inclusive educational resources. For policymakers, a data-driven approach is essential to crafting data-based AI policies that account for real user patterns and concerns, rather than relying solely on qualitative methods or anecdotal evidence. For academia, the integration of fact-checking protocols is crucial to combat misinformation and uphold academic integrity, ensuring that LLMs are used as tools to enhance, and not replace, critical thinking and scientific reasoning. This study serves as a call to action for all AI stakeholders to collaboratively shape a future where AI tools empower us and are responsibly integrated into society.



\section{Research Ethics and Social Impacts}\label{sec:ethics}
\subsection{Ethical considerations}

The study received approval from the Institutional Review Board (IRB), ensuring compliance with ethical standards for human subjects research. Participants were provided with a detailed informed consent form outlining the study and data usage. Personally identifiable information (e.g., email addresses) was collected for a data cleaning process and information about their identity (e.g., gender, sex, race, etc.) was not collected. Unique identifiers were used to protect participant identities rather than using their name. Participation was entirely voluntary. For this research, the survey was conducted exclusively in English, which may have limited participation from non-English speakers. This limitation is acknowledged, and future studies will aim to include multilingual options to ensure broader representation.

\subsection{Positionality}
The authors are researchers in STEM fields and the study was conducted at an engineering-centric R1 research university. Our backgrounds could have influenced the research frame as well as shaped the participant pool and results. For example, the institution’s focus on STEM and engineering may have influenced the high-value participants placed on understanding the technical processes behind LLMs. We recognize that more exposure to multiple types of large language models may have influenced the replies. We have taken steps to mitigate bias by consulting with experts in user studies and IT policy during the qualitative analysis phase.

The authors are researchers in STEM fields, and the study was conducted at an engineering-centric R1 research university. Our academic backgrounds and institutional context may have influenced the framing of the research, the composition of the participant pool, and the interpretation of results. For instance, the institution’s strong emphasis on STEM and engineering likely contributed to participants placing high value on understanding the technical processes underlying LLMs. Additionally, the level of exposure to various types of LLMs and technology-optimistic culture may have shaped their responses. To address potential biases, we consulted experts in user studies and IT policy during the qualitative analysis phase, ensuring a more balanced and rigorous approach to data interpretation.

\subsection{Adverse and unintended impacts}
While this study aims to inform ethical and effective LLM policies in academic environments, there is a risk that the findings could be misinterpreted or misapplied. For instance, while fact-checking mechanisms are essential for ensuring accuracy and reliability, an overemphasis on them might lead to overly restrictive policies that inadvertently stifle innovation, particularly in contexts requiring creative or exploratory text generation. Striking a balance between rigorous fact-checking and fostering creative freedom is critical to avoid unintended consequences. Additionally, the study focuses exclusively on academic settings, which may limit its applicability to other sectors, such as government or industry, where the use cases and regulatory frameworks for LLMs may differ significantly.


\clearpage
\bibliographystyle{ACM-Reference-Format}
\bibliography{refs}


\clearpage
\appendix

\section{Appendix}\label{sec:appendix}
\subsection{Figure A1: LLMs are being utilized for a wide range of activities}

\begin{figure*}[htbp]
  \centering
  \includegraphics[width=0.9\textwidth]{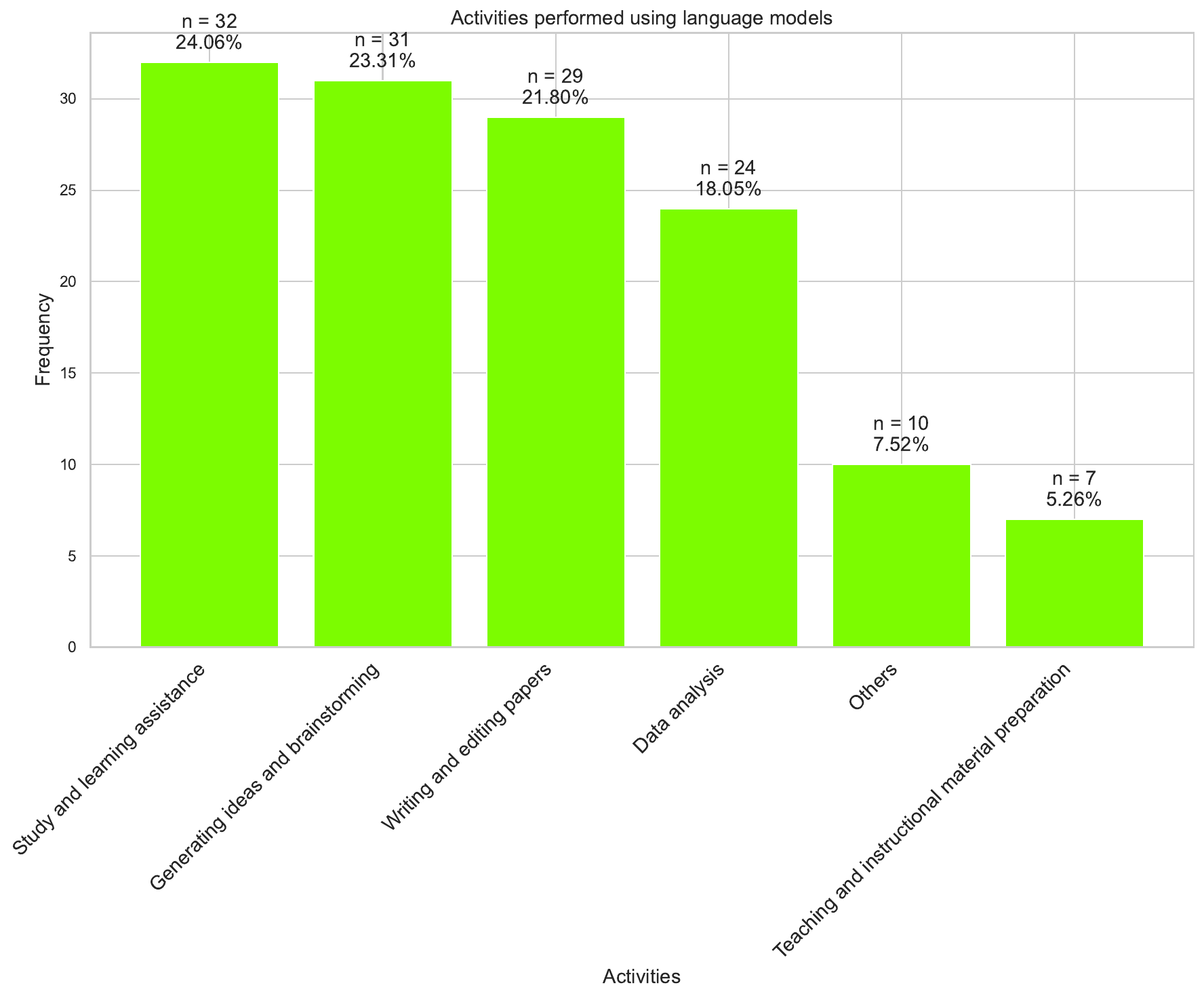}
  \caption{This figure illustrates the diverse applications of LLMs across various tasks, from research and learning to data analysis, highlighting their versatility.}
  \label{fig:purposes}
\end{figure*}

Figure \ref{fig:purposes} illustrates diverse tasks that people work on using LLMS for research, showing these tools' multifaceted role in academia. From brainstorming and generating ideas to assisting with studies and editing written material, LLMs were leveraged across various stages of academic work. The utility of these models extends into more specialized tasks as well, such as data analysis. This variety suggests that LLMs have become a staple in the toolkit of people in academia, supporting a wide spectrum of intellectual activities. This part of the survey allowed multiple choices because individuals often utilize LLMs for various tasks.


\subsection{Figure A2: Predominance of ChatGPT in LLM Usage}

\begin{figure*}[htbp]
  \centering
  \includegraphics[width=0.9\textwidth]{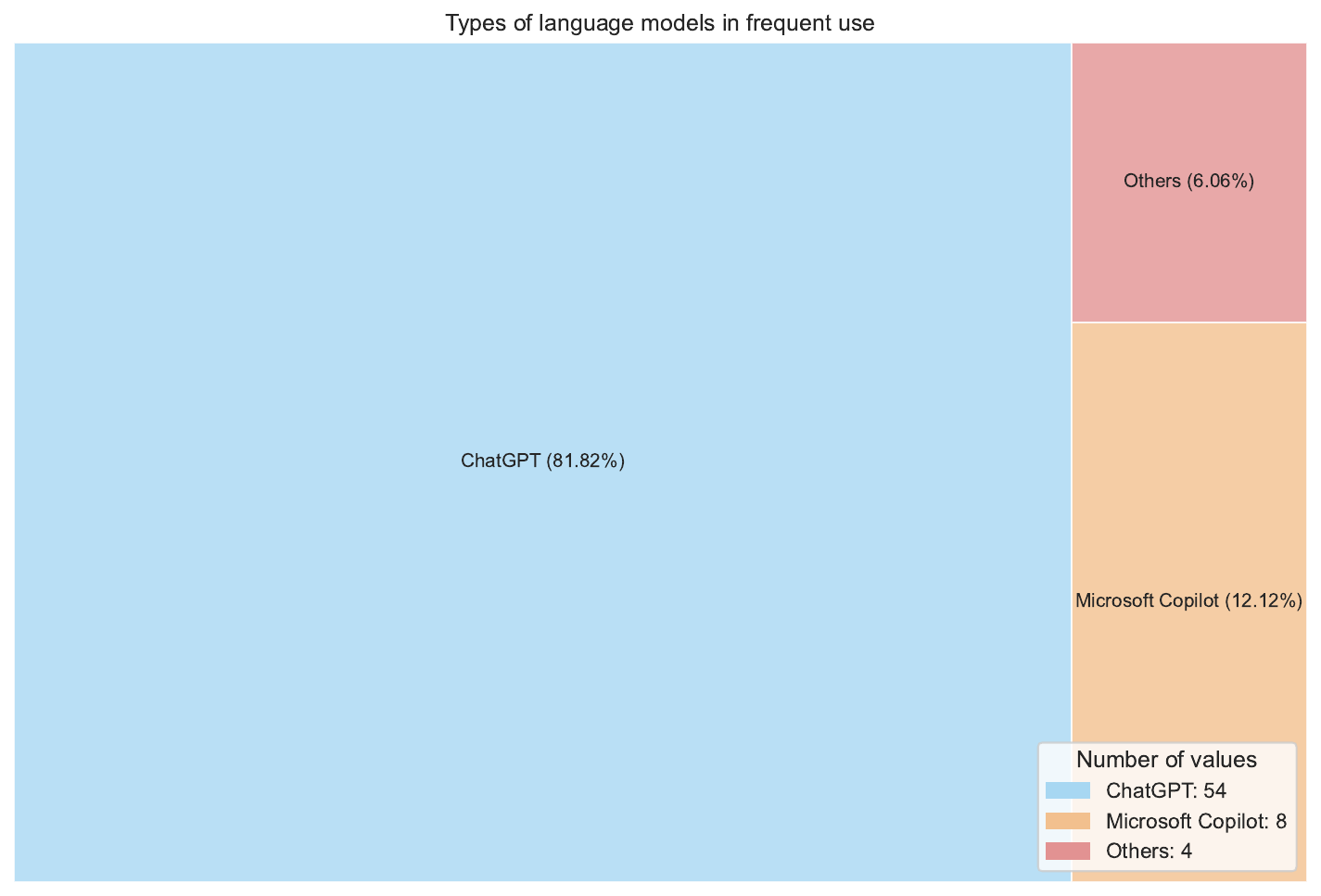}
  \caption{ChatGPT dominates as the most widely used language model, with a significant preference towards OpenAI's tool. Microsoft's Copilot follows as the second most utilized model.}
  \label{fig:models}
\end{figure*}

Considering the responses gathered, it becomes evident that ChatGPT holds a significant position among AI users. Figure \ref{fig:models} shows the proportion of the types of LLMs that the respondents use. OpenAI's ChatGPT took the greatest portion, with 81.82\% respondents indicating its use when we ask about the mostly engaged LM. In comparison, Microsoft Copilot garnered 12.12\% of mentions, while around 6\% of individuals cited other LLMs that are not famous. 

Participants were asked to select one language model they primarily use from the following options: ChatGPT, Microsoft Copilot, Google Bard, a fine-tuned model specific to their field, or others. The replies were collected from people who use LLMs. The detailed model names for ``Others'' were not addressed because of the absence of the prompt to type the reply.

ChatGPT, a prominent LLM developed by OpenAI, can be considered a ``metonymy'' of the LLMs. Metonymy is a way of speech where one word or phrase is substituted for another with which it is closely associated. For example, when the news said ``The White House announced a new policy today,'' \textit{The White House} is used as a metonymy for the executive branch of the United States government. The physical building known as the White House is closely associated with the executive branch and is often used to represent the actions and decisions of the President and their administration. In the case of ChatGPT, it could symbolize a broader category of models that utilize similar architectures and methodologies for text generation and understanding. When individuals mention using AI text generators, they can refer not only to a specific developed model but also to a wider spectrum of LLMs built upon similar methods.

\subsection{Figure A3: Participants' Roles in the Study}\label{fig:roles_school}

\begin{figure}[htbp]
    \centering
    \includegraphics[width=3.3in]{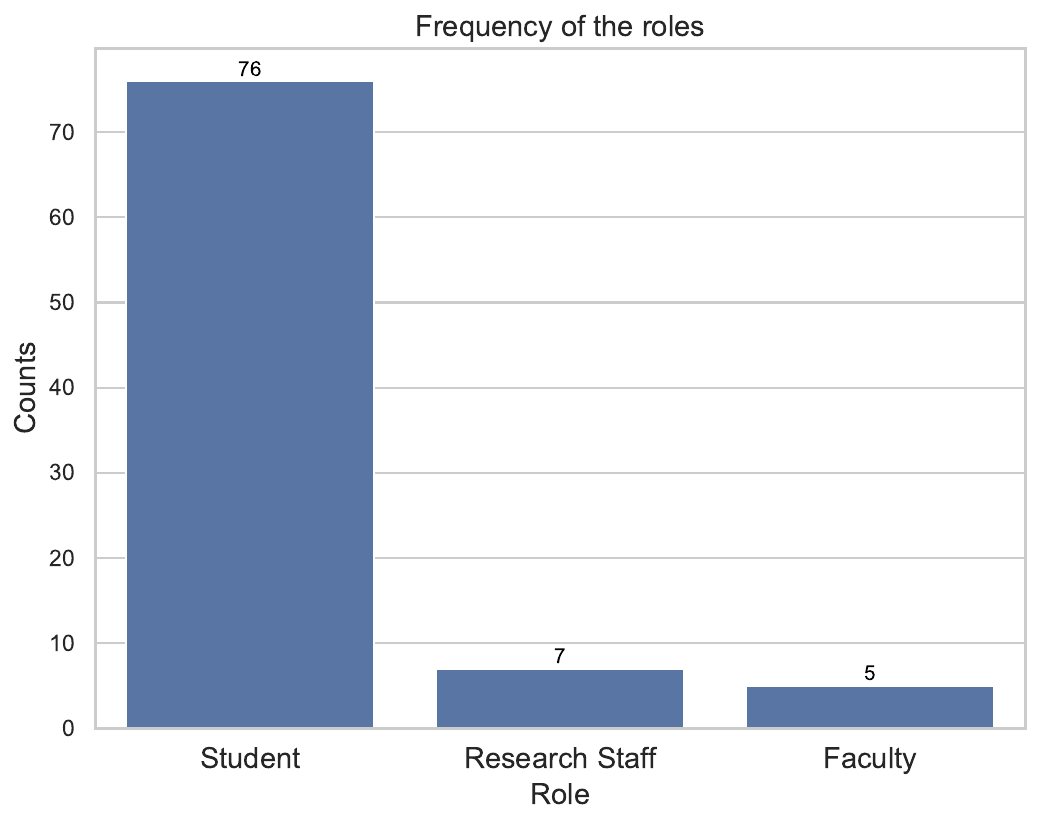}
    \caption{Participation breakdown: 76 students, 12 non-students (7 research staff and 5 faculty members) contributed to this research study.}
    \label{fig:roles}
\end{figure}

Figure \ref{fig:roles} shows the distribution of survey respondents across three categories: Students, Research Staff, and Faculty. The data reveals that the vast majority of respondents were students, accounting for 76 responses and 12 non-students were participated, research staff and 5 faculty members.

In this research, the distinction between undergraduate and graduate participants was not specifically considered, particularly since some people in the school consider some master's programs as an extension of undergraduate study and some master's student would reply that they are 5th year undergraduate student. Non-academic staff were excluded from the study population to ensure that the focus remained on the use of language models in academic activities, aligning with the study's aim to explore their role in educational and research contexts rather than across all roles within the institution.

\subsection{Figure A4: Importance of Understanding the Process Behind LM}\label{appendix:understanding_process}

\begin{figure}[htbp]
    \centering
    \includegraphics[width=3.3in]{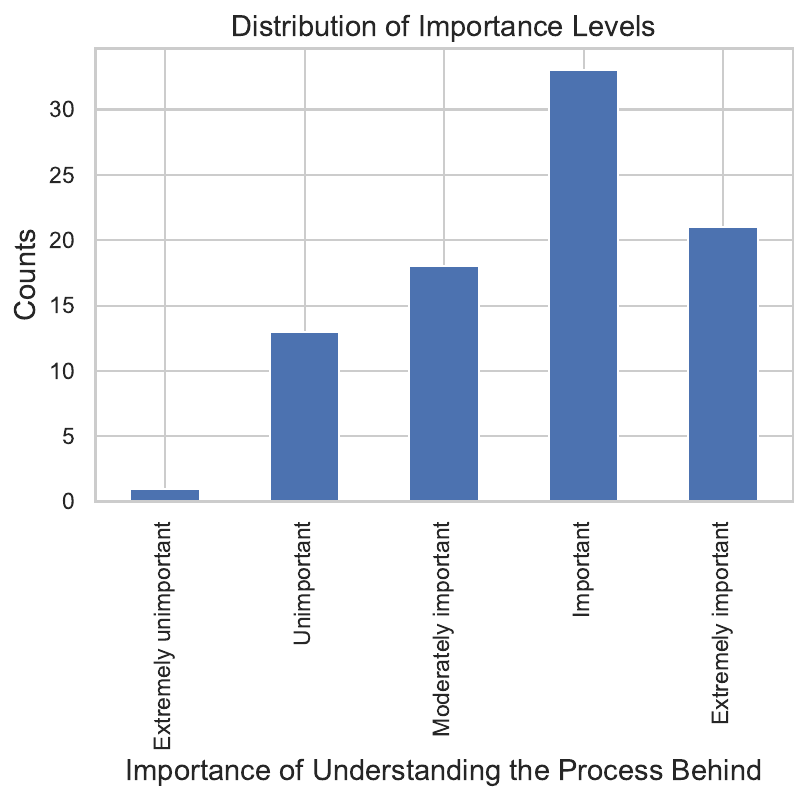}
    \caption{The majority of respondents replied that understanding the process behind LM is important. This shows the characteristics of the school which the study was conducted.}
    \label{fig:processbehind}
\end{figure}

Figure \ref{fig:processbehind} illustrates the distribution of respondents' opinions regarding the importance of understanding the underlying processes behind LLMs. As mentioned in the ``Limitations and Future Work'', this result shows the characteristics of the study population that considers the engineering process crucially. Regarding the importance of understanding the AI generation process, there is an emphasis on ``Important'' or ``Extremely important.'' The distribution of responses was as follows: 33 participants found the process ``Important,'' 21 rated it ``Extremely important,'' 18 considered it ``Moderately important,'' 13 deemed it ``Unimportant,'' and 1 viewed it as ``Extremely unimportant.'' This implies that the school, being a technology-oriented institution, values a deep understanding of its underlying processes and it shows the characteristics of the institution where the study was conducted.

\newpage
\section{Appendix B: Additional Information on Dataset and Experimental Process}

\subsection{Datasets}
\begin{itemize}
    \item This paper is based on a single dataset, derived from a one-time survey conducted at a specific institution. The nature of this study required the dataset to be collected in a singular instance to maintain the integrity and validity of the experimental conditions, particularly concerning the exposure of participants to the question prompts, because repeating the survey could compromise these factors and potentially bias the results.
\end{itemize}

\subsection{Computational Experiments}
\begin{itemize}
    \item This paper includes computational experiments conducted using Python. While we intend to make the code publicly available, there are privacy considerations that prevent us from doing so immediately. Specifically, some parts of the code refer to the institution where the study was conducted, such as in the filtering process the code explicitly addresses the institution’s email domain for effective filtering (e.g. @[institution].edu). Additionally, the dataset includes sensitive typed-in responses from participants that require careful processing for their privacy as well as contains language that may be offensive to others. We are committed to adhering to the Institutional Review Board (IRB) policies and complying with the privacy terms agreed upon by participants at the start of the study. As such, any disclosure of the code will be managed in a manner that complies with these policy considerations.
\end{itemize}

\newpage
\section{Appendix C: Details of Data analysis}\label{appendix:quantatitive_analysis}
\subsection{Frequency analysis}
First, this study observed the distribution of responses regarding LM usage, specifically measuring the proportion of participants actively using LLMs. The study used a standard histogram frequency analysis using five bins corresponding to the five levels of the Likert scale. This analysis aimed to assess whether the majority of people are engaged with these tools and to measure the type of engagement based on use time.

\subsection{Point-biserial correlation coefficient}\label{appendix:point-biserial} 

Secondly, we investigate the relationship between LM adoption and trust levels as measured by a correlation analysis. Since adoption is a binary categorical variable (non-users v. users) and trust levels are ordinal data, a point-biserial correlation analysis was deemed the most appropriate method. We use the function below \cite{cureton1956rank},

\begin{equation}
r_{pb} = \frac{M_1 - M_0}{s_{n-1}} \sqrt{\frac{n_1 n_0}{n(n-1)}},
\end{equation}

where \( r_{pb} \) represents the point-biserial correlation coefficient, \( \overline{X_1} \) and \( \overline{X_0} \) are the mean trust levels of users and non-users respectively, \(s_{n-1} \) is the sample standard deviation of trust scores, and \( n_1 \) and \( n_0 \) denote the number of users and non-users; \( n \) is the sum of \( n_1 \) and \( n_0 \).

A subsequent t-test \cite{kornbrot2014point} was conducted to \( r_{pb} \) to evaluate the statistical significance of the observed correlation coefficient:

\begin{equation}
t = \frac{r_{pb} \sqrt{n-2}}{\sqrt{1-r_{pb}^2}}
\end{equation}

The t-value was calculated using the derived \( r_{pb} \) and \( n \).


\subsection{Kendall's tau}\label{appendix:kendall-tau} 
Having studied the binary adoption of LLMs, we now go one step further to analyze the ordinal association between self-reported use time and trust levels. In contrast to the model adoption being binary, use time is ordinary. The analysis using Kendall's tau \cite{kendall1938new} measured the ordinal association between these two variables, assuming a non-parametric distribution:


\begin{equation}
\tau = \frac{2(C - D)}{n(n-1)}
\end{equation}

Kendall's tau (\(\tau\)) accounts for the similarity between the proportions of concordant and discordant pairs. A pair of observations is considered discordant (\(D\)) if, when comparing two observations, the relative order of the two variables differs. Specifically, if one variable increases while the other decreases, or vice versa, the pair is \(D\). In this case, if the usage time rank increases but the trust level rank decreases, \(D\). Otherwise, the pair is concordant (\(C\)).
The \(\tau\), provides a measure of the strength and direction of association between the two ordinal variables. The analysis was further validated using the p-value for Kendall's tau, corresponding to a Z-score \cite{schober2021statistics}, calculated through the normal approximation method:

\begin{equation}
Z = \frac{\tau}{\sqrt{\frac{2(2n+5)}{9n(n-1)}}}
\end{equation}

Additionally, the Locally Weighted Scatterplot Smoothing (Lowess) regression technique was employed to fit a smooth curve to the data in \ref{fig:correlation2}, which assumes a non-linear relationship between variables \cite{cleveland1988locally}.

\subsection{Ranking}\label{appendix:ranking}
Our next analysis aims to study the issue to prioritize among five core issues mentioned in \textit{RQ4}: fact-checking, editing, ethical
decision-making, restricting, or disclosing. We used a score-based ranking method using a poll based on a Likert scale of 1 to 5 to identify the primary concern among the major topics.

\subsection{Qualitative analysis}\label{appendix:qualitative_analysis}

A qualitative analysis of users' written rationales was conducted to gain deeper insights into their perspectives and reasoning. The first author engaged in a two-month consultation with experts in user study and IT policy research and thoroughly reviewed their responses. Replies were briefly summarized, and these summaries were then synthesized into a document by each core issue: \textit{fact-checking, editing, ethical decision-making, restricting, or disclosing}, incorporating direct quotations as evidentiary support for the content analysis. The human review focused on identifying key themes, contextual nuances, and implicit meanings in the responses, ensuring a comprehensive understanding of the participants' perspectives. Direct quotations from participants were used to support the findings and for the conclusions drawn.

\section{Appendix D: Qualitative analysis and summaries}\label{appendix:opinion_summaries}
\subsection{Fact-checking}
Respondents who cast votes for each policy highlighted the need to uphold the highest standards of accuracy and scientific integrity, replying to mandatory fact-checking as a crucial barrier against the spread of misinformation. They also highlighted a preventive measure to discourage misuse, particularly in academic settings, ensuring users do not overly rely on AI without critical reasoning and investigating.

In the written rationales for their selections, respondents (denoted as “P”) stressed the importance of fact-checking to prevent misinformation and to secure scientific integrity, saying “\textit{a strict rule of human verification before publication is essential}” (P76) and noted needs for guidelines or policies to "\textit{[m]aintain the highest scientific standards" (P31) as scientific standard of academia impacts to outside of academic community as well as related to future development of society. People who work in an academic environment "should not blindly trust the output}" (P46) since the generated information is far from orthodox scientific reasoning and scientists are accountable for the information they developed. People stated we need to maintain a high academic standard to organize a constructive academic society rather than let generated words from AI occupy the mainstream of science. 

The P58 noted that he does not trust information from LLMs: "\textit{It's similar to not trusting material for Wikipedia.}" Wikipedia is not considered a trustworthy resource to cite because its publication process is not credible and the content itself can be changed. That is to say, those anonymous authors are not responsible for the accuracy, bias, or vandalism of the content. In contrast to Wikipedia, academic publications are authored by identifiable individuals who take responsibility for the knowledge they present. These authors are expected to produce high-quality, accurate, and appropriate information. The onus is on users to verify this information themselves, rather than relying on AI to handle these essential verification tasks. 

There were two contrasting viewpoints from the rationales provided by the respondents: strong regulation of AI tools and minimal regulation. On the side for the strict regulation, P49 expressed needs of strong regulation of the usage: \textit{"students won’t care unless there is an actual rule governing that they cannot use AI without fact checking"}. In contrast, P41 noted that LM usage is pervasive and "\textit{these technologies are here to stay}" and argued for not having a regulation. 

In the survey, 61.11\% of respondents noted the needs for human evaluation, especially in fields that needs meticulous review, while arguing minimal regulation. This is a procedural requirement that we need to bring up explicitly to lead the action of the authors. To put it another way, we need reasonable enough boundaries not to make an excessive regulation that could frustrate the better usage of AI tools. As P92 noted, \textit{"it is important to use generative ai"} and it has the potential to enhance the reasoning and development process. That is to say, we need to figure out a medium between full regulation and no regulation to make appropriate range of the quality control in order to "\textit{maintain the highest scientific standards}" (P31) as well as to take advantage of the AI tools as P41 and P92 signified. Here we could pose strict regulations in fields that need accurate information like medical research. In contrast, we could allow more flexibility in creative domains such as science fiction writing.

Also, voters highlighted double-checking before the public disclosure. That is to say, the publication process that share information with people is considered importantly. 38.89\% of respondents noted procedural protocol. That is to say, respondents expressed the need for using AI tools in the generation process but claimed needs for double-checking before the public disclosure that is related to dissemination of knowledge generated by the AIs. P 92 noted that we should integrate verification process to encourage people to work on \textit{"double checking it is fine"} to use the generated knowledge because of its nature of stochastic processing. P76 also noted that "\textit{a strict rule of human verification before publication is essential}." It is important not "\textit{[lose] critical thinking skill[s]/ fact checking skills}" (P14) in all stages of knowledge generation. 

We also need to ensure that we do not have an unnecessarily burdensome policy. In respondents' opinions related to fact-checking, a standardized policy that works for all cases was disfavored because the universal ruling could miss potential benefits in specific fields though it can provide simplified regulations. Excluding one-size-fits-all approaches and unrealistic compliance expectations from the policy framework is considered essential to prevent it from becoming overly burdensome or impractical. Approximately 27.78\% of the rationales noted concerns about one-size-fits-all approaches. By avoiding overly broad policies, we acknowledge that different content areas carry varying levels of risk and therefore require tailored guidelines. Participants expressed concerns that overly restrictive policies could hinder productivity and the beneficial use of AI tools. As one respondent pointed out, \textit{"AI is a tool that can be helpful"} (P46) in many cases, emphasizing the importance of leveraging AI's advantages without being encumbered by excessive regulations.

Furthermore, adaptability is crucial in the face of rapidly evolving technology. Overly prescriptive policies risk becoming obsolete or stifling innovation. Approximately 33.34\% noted concerns about unrealistic compliance expectations, and P 41 noted that \textit{"these technologies are here to stay"}. By excluding inflexible and overly broad requirements, we ensure that the policy can adjust to new developments in AI and support users in adopting beneficial technologies while maintaining necessary safeguards.

\subsection{Editing}
Editing, which includes organizing, paraphrasing, and writing one's ideas and words, is a crucial reasoning task we need to preserve.

Respondents emphasized the need to prevent the deterioration of writing process and the creation of a homogeneous body of text lacking individual creativity: P47 noted "\textit{We risk the utter deterioration of people’s writing ability, and a boring world where everything is written to sound exactly the same, with no written communication skills developed}." Additionally, there was a preference for a proactive editorial approach that aligns with established scholarly norms. P106 noted that "\textit{[users are] fully responsible for any text you publish as though you had written it yourself.}, necessitating responsible writing rather than copying and pasting. Some respondents also expressed a need to avoid AI-generated content stressing the importance of preserving the integrity and originality of scholarly work to "\textit{maintain high scientific standards} (P29)." Also, the need for education to prevent academic plagiarism is pronounced (P66). Additionally, P90 emphasized the importance of developing an argument in the paper: \textit{"the crux of the argument or fact is the important part that should not be generated by ai,"} expressing the need for humans working in crafting essential arguments and ensuring the integrity and originality of the content.

However, some people noted that a more relaxed approach could be beneficial. Advocates of a laissez-faire policy argue for greater flexibility and freedom in content creation. For instance, P80 stated, \textit{"[l]et people write and say what they need to,"} emphasizing individual autonomy in writing. Similarly, P106 expressed support for the integration of AI tools, asserting, \textit{"I believe it is fine to use text generated by AI,"} highlighting the potential for AI to assist academic work. 

In analyzing the respondents' rationales, two primary themes emerged regarding the editing of AI-generated content: the need for detailed editorial criteria to enhance content quality and the importance of avoiding uniform regulation across domains and rigid publication barriers. P74 emphasized responsible editing by stating, "\textit{If you use it [...], you are responsible for those [content]}," highlighting that authors are ultimately accountable for the content. Although some people argued for the free usage policy for freedom in a generation, people highlighted the importance of overall quality as well as integrity in academia. 

In the survey, 38.46\% of respondents advocated detailed editorial review criteria, focusing on explicit editorial criteria for improving content quality, such as avoiding copying wording and framing from the LLMs. P29 pronounced the importance of ``\textit{maintaining high scientific standards}''. P90 noted ``\textit{[a lot] of text is not novel}'' among generated content and ``\textit{novel thoughts should be properly highlighted}'', and stated that even though collaboration with AI tools can greatly enhance productivity and bring up some novelty, researchers themselves need to work on the crucial part of the paper, like the main argument, rather than delegating all generation process to the language models. People noted the needs for proactive quality control to impose minimal requirements to authors to edit through AI-generated output for high scientific standards, which is related academic integrity, and novelty.

Also, 23.07\% of people expressed the need for the exclusion of the rigid publication barrier and argued the need to avoid policies that impose excessive editorial burdens. P54 acknowledged difficulty in implementing overly strict verification processes and P116 highlighted the trade-offs between supervisory effort and time costs: \textit{"There is a trade-off to be made in human's supervisory role and responsibility vs. time cost into checking everything through the editorial process"}. Also, P106 noted \textit{"I believe it is fine to use text generated by AI}" and advocated for user autonomy and responsibility rather than imposing the publication barriers.

\subsection{Ethical decision-making}
The process of ethical decision-making, which includes the evaluation and verification of AI-generated content and its applications, is of utmost importance in guaranteeing the appropriate use of AI-generated content. We should refrain from using unethical content that includes racism, gender bias, and discrimination. Despite some sentences could be true statements, their usage needs profound consideration.

Ethical judgment regarding AI-generated content is deemed critically important by many stakeholders. Concerns are raised about the potential for language models to produce inappropriate or biased phrases, such as content that inappropriately targets specific groups (e.g., group \texttt{\(G_1\)} tends to be good in task \texttt{\(W_1\)} but bad in \texttt{\(W_2\)}). For instance, P85 noted: \textit{"Humans should design their own society and not let virtual agents tell us what is important,"} claiming the necessity for human oversight in determining the appropriateness of content. Authors are arbiters of content quality, diligently excising any discriminatory or unethical statements, such as those advocating for gender-based considerations in employment.

Furthermore, the societal context that the output text could potentially used or intentionally disclosed should be carefully considered by authors. That is to say, the authors are expected to work on crucial evaluation and moral reasoning. P85 asserted human responsibility through moral judgment by saying \textit{"[h]umans should not rely too much on artificial intelligence. What people think is also very important."} This perspective is reinforced by P116, who remarked: \textit{"Human's role in final decision making is central,"} underscoring the humane role that plays in ensuring ethical standards of scholarly work. P66 expressed worries by stating: \textit{"I don't think it should be used to generate material that is responsible for communicating with other people, i.e., writing papers."}

In contrast, others like P90 acknowledged the inevitability of AI-generated text, stating, \textit{"[j]ust accept that [the] time has come [when] AI-generated text will be there,"} noting that embracing AI's presence is unavoidable.

In the research, 57.14\% of participants expressed the need for addressing sensitive features in AI applications. Sensitive features are attributes like race, gender, and other personal characteristics that relate to vulnerable populations and can be subject to bias or discrimination. Especially, P47 noted: \textit{"We’re releasing and using tools that are already being used to wreak havoc in society (election and war disinformation, deepfake pornography, synthetic voice extortion schemes)"} and P82 asserted the need for human consideration and interruption to address these issues, saying \textit{"Humans should design their own society and not let virtual agents tell us what is important"}. P85 pointed out the importance of the independent moral reasoning by noting \textit{"[h]umans should not rely too much on artificial intelligence. What people think is also very important."} To put it another way, we need to promote our independent thinking that does not rely on generative tools and can tell what can be addressed or not as well as what to tell or not. Also, P116 highlighted the crucial role of human final decision-making: \textit{"Human's role in final decision-making is central."} We also need to consider that the unique characteristics of the mostly voted policy notes the needs for the explicit notations for ethical compliance.

On the contrary, 42.85\% of people expressed the need for the exclusion of the prescriptive regulation on ethics. P80 emphasized that AI tools should not be regulated more strictly or leniently than other tools and agents, stating, \textit{“[e]valuate AI tools with the same criteria as we do non-AI works and opinions.”} P90 noted \textit{"it is choosing the middle ground. Banning the tech or highly regulating it is no solution. Just accept that the time has come when AI-generated text will be there. Just pass it through the same pair of scrutinizing eyes as human-generated text. Banning it is like killing a baby"}, expressing a more flexible usage policy for AI tools as multiple respondents keep signified.

\subsection{Restricting}
The majority of people consider policy around restricting is important and we should have policy to govern the language model usage in academia.

The reasons behind the restriction policy selections emphasize the need for institutional backing to guide AI usage, endorse the integration of AI for enhancing learning efficiency and creativity, and advocate for an appropriate guideline to interact with technological advancements in academic settings. 

Stakeholders assert that institutional support is crucial to provide clear guidelines and frameworks for the responsible use of AI. For instance, P67 asserted, "Instructors should be backed by the institute, otherwise it gives students too much leeway,"emphasizing the need for developing institutional policies to support both coursework and research by establishing evidentiary guidelines tailored to individual educational and research contexts. This aims to prevent misuse and ensure that AI tools are utilized effectively and appropriately within the academic setting. Additionally, P85 emphasized the importance of fostering independent thought among students through partial regulation in using AI tools by stating, "[s]tudents should cultivate their own thinking," suggesting that while AI can be a valuable tool, it should not replace the development of critical thinking skills.

Some people advocate for greater flexibility and openness in the use of AI, arguing that strict institutional controls may stifle creativity and hinder the potential benefits of AI integration. P116 remarked: "allow flexibility depending on specific needs in the academic setting," and noted that policies should be adaptable to accommodate diverse academic requirements and individual preferences, implying a nuance that blue law should not established unless it certainly aids the goal of research or education. P80 took a more enthusiastic stance toward AI integration, stating,``Let AI be AI! If students want to have AI write their paper and be creative, why isn't that a good thing? If people want to gain knowledge and information in the fastest and most efficient way possible, it would be deplorable to stop them from doing so,'' expressing the belief that AI can significantly enhance learning and creativity.

Considering that the majority of people replied that a policy that addresses restriction is important and strongly important, this issue needs significant consideration. But some people argue for \texttt{Laissez Faire policy} (i.e. no interruption policy or no regulation) for the freedom to use technologies despite people on partial regulation worries about over-reliance. 

Approximately half of the respondents proactively advocated for a usage-oriented policy that encourages active use of AI tools rather than banning them. P12 emphasized the importance of the synergy that can be achieved when AI is used wisely and strategically. Supporting this view, P80 argued for a free usage policy, stating: \textit{"Let AI be AI! If students want to have AI write their paper and be creative, why isn't that a good thing? If people want to gain knowledge and information in the fastest and most efficient way possible, it would be deplorable to stop them from doing so."} This perspective is echoed by P82, who asserted, \textit{"Research institutions should be open-minded on technical breakthroughs,"} advocating against restrictive policies and for openness in the use of AI tools. Consequently, it is imperative to adopt a more permissive stance towards the utilization of AI tools, rather than imposing restrictions. That is to say, the usage policy should be oriented to policy for full allowance or partially allowance rather than restrict its usage. There exists potential for enhanced application and synergy, thus the usage policy should be flexible enough by considering the partial employment although restriction is deemed optimal.

One-third of the respondents emphasized the importance of providing reasonable rationales when implementing regulations. They suggested that while the institution may attempt to regulate the usage of AI tools, any such regulations should be supported by sound justifications rather than merely imposing restrictive rules without explanation. P67 stated: \textit{"I think a blanket statement is the wisest way to go instead of delegating that decision to instructors. Instructors should be backed by the institute; otherwise, it gives students too much leeway."} Also, P116 noted that it is \textit{"[h]elpful to allow flexibility depending on specific needs in the academic setting."}

Approximately 35\% of respondents emphasized the need to avoid banning AI tools without a thorough review. Since implementing such a significant regulation requires justification, the regulatory body must provide clear rationales for any decision to ban AI tools. While college students should indeed be encouraged to develop their thinking through writing and critical analysis, as P85 mentioned: \textit{"[c]ollege students should cultivate their own thinking,"} we must also consider the effectiveness of writing as a tool for improving reasoning. Moreover, it is important to acknowledge that AI tools have become an integral part of research. Especially, P116 noted that AI tool usage policies should not make \textit{"one seems to fool oneself for avoiding new technology and will one day be left behind by the new generation."}

\subsection{Disclosing}
The disclosure of details regarding AI usage, including the specific models utilized and the tasks assigned to these models, is considered an important issue as well.

The rationales for the selection related to disclosing policy highlighted the importance of proper citations and acknowledgments. P60 stated, \textit{"If a LLM is used, it needs to be fully and properly documented in the same way that computational science researchers acknowledge their methods, software, etc."}  He insisted the stance that the AI tool is a kind of research tools like other software and models that we need to cite in research and we need to keep the same level of transparency by citing them. This ensures that the use of AI is openly acknowledged, allowing for reproducibility and accountability in research. P87 also articulated this perspective:\textit{"It’s a tool."} P116 stressed that reproducibility lies at the heart of research: \textit{``"Reproducibility that is central to research should be upheld."''}. Detailed attributes, especially about the LM tools that was used for the work, is important because the generation process of each LM is different. Description about the detailed model, used prompt, and rationales for the selection could be helpful for other research by letting other researchers to use that method.  

Reliability of the research work is issued by many people. P90 emphasized \textit{"[s]hould not rely entirely on artificial intelligence,"} advocating referring to parts that are generated by LLMs to make the work distinguishable as well as preventing the work 100\% generated by LLMs. This perspective is complemented by P116, who asserted, \textit{"[r]eproducibility that is central to research should be upheld,"} reinforcing the necessity of maintaining reliable and authoritative research standards through comprehensive notification and citation practices. These viewpoints collectively stress that while AI can be a valuable asset, it should not replace fundamental human responsibilities.

Conversely, some stakeholders argue against imposing stringent requirements for the acknowledgment. P80 expressed a more flexible stance by suggesting that rigid regulations may hinder the practical and innovative application of AI in various academic contexts: \textit{"what can an AI output that humans couldn't, given sufficient time?"}.

The need to include a policy that requires users to articulate one's rationale for employing AI tools is underscored by 42.85\% of participants. P60, for example, insists on providing a thorough documentation of AI usage in a manner analogous to citing traditional computational methods or specialized software by saying: \textit{"If an LLM is used, it needs to be fully and properly documented in the same way that computational science researchers acknowledge their methods, software, etc."} Such a stance treat LLMs as integral components of the research apparatus, worthy to acknowledge the rationale for the selection and usage like the employment of specific equipment or hardware in other research.

On the other hand, people noted that we need to refrain from posing an excessive requirements for the attribution for every minor instance to avoid unnecessary administrative burdens and procedural rigidity. Approximately 45\% of respondents shared opinions that disagree with the overly excessive attribution policy. As P80 suggests, guidance rather than stringent directives is more conducive: \textit{``Guidance is the best—not requirements for AI tools.''}. Also, P66 noted that excluding the restrictive attribution policy could help maintain a productive balance.





\end{document}